\documentclass{edm_template}

\usepackage{booktabs} 
\usepackage{ftnxtra}
\usepackage{fnpos}
\usepackage{lmodern}
\usepackage{amsmath}
\usepackage{amsfonts}
\usepackage{array}
\usepackage[caption=false,font=footnotesize]{subfig}
\usepackage{stfloats}
\usepackage{url}
\usepackage{multirow}
\usepackage[normalem]{ulem}
\usepackage{threeparttable}
\usepackage{hyperref}
\usepackage{booktabs}
\usepackage{graphicx}
\usepackage{subfig}
\usepackage{tikz}
\usepackage{amsmath}
\usepackage{mathtools}
\usepackage{amssymb}
\usepackage{color}
\usepackage{longtable}
\usepackage{array}

\newcommand{\method}{NAK}

\newcommand{\q}{\mbox{$\mathbf{q}$}}
\newcommand{\p}{\mbox{$\mathbf{p}$}}
\newcommand{\x}{\mbox{$\mathbf{x}$}}

\newcommand{\req}{\mbox{$\mathbf{r}$}}
\newcommand{\ks}{\mbox{$\mathbf{k}$}}

\newcommand{\term}{\mbox{$\mathcal{T}$}}

\newcommand{\G}{\mbox{$\mathbf{G}$}}

\newcommand{\etal}{{\em et al.}}
\newcommand{\ul}{\underline}
\newcommand{\red}{\textcolor{red}}

\graphicspath{{./fig/}}
\DeclareGraphicsExtensions{.png}

\begin{document}

\title{Sparse Neural Attentive Knowledge-based Models\\for Grade Prediction}

\numberofauthors{2}
\author{
\alignauthor
Sara Morsy\\
       \affaddr{Department of Computer Science \\\& Engineering}\\
       \affaddr{University of Minnesota}\\
       \email{morsy@cs.umn.edu}
\alignauthor
George Karypis\\
       \affaddr{Department of Computer Science \\\& Engineering}\\
       \affaddr{University of Minnesota}\\
       \email{karypis@cs.umn.edu}
     }

\maketitle
\begin{abstract}
  Grade prediction for future courses not yet taken by students is important as
  it can help them and their advisers during the process of course selection as
  well as for designing personalized degree plans and modifying them based on
  their performance. One of the successful approaches for accurately predicting
  a student's grades in future courses is Cumulative Knowledge-based Regression
  Models (CKRM). CKRM learns shallow linear models that predict a student's
  grades as the similarity between his/her knowledge state and the target
  course. A student's knowledge state is built by linearly accumulating the
  learned provided knowledge components of the courses he/she has taken in the
  past, weighted by his/her grades in them. However, not all the prior courses
  contribute equally to the target course. In this paper, we propose a novel
  Neural Attentive Knowledge-based model (\method) that learns the importance
  of each historical course in predicting the grade of a target
  course. Compared to CKRM and other competing approaches, our experiments on a
  large real-world dataset consisting of $\sim$1.5 grades show the
  effectiveness of the proposed \method\ model in accurately predicting the
  students' grades. Moreover, the attention weights learned by the model can be
  helpful in better designing their degree plans.
\end{abstract}

%


\section{Introduction}

The average six-year graduation rate across four-year higher-education
institutions has been around 59\% over the past 15
years~\cite{kena2016condition,braxton2011understanding}, while less than half
of college graduates finish within four
years~\cite{braxton2011understanding}. These statistics pose challenges in
terms of workforce development, economic activity and national
productivity. This has resulted in a critical need for analyzing the available
data about past students in order to provide actionable insights to improve
college student graduation and retention rates.

One approach for improving graduation and retention rates is to help students
make more informed decisions about selecting the courses they register for in
each term, such that the knowledge they have acquired in the past would prepare
them to succeed in the next-term enrolled courses. Polyzou
\etal~\cite{polyzou2016grade} proposed course-specific linear models that learn
the importance (or weight) or each previously-taken term towards accurately
predicting the grade in a future course. One limitation of this approach is
that in order to make accurate predictions, the model needs to have sufficient
training data for each (prior, target) pair. Morsy
\etal~\cite{morsy2017cumulative} developed Cumulative Knowledge-based
Regression Models (CKRM) that also build on the idea of accumulating knowledge
over time. CKRM predicts a student's grades as the similarity between his/her
knowledge state and the target course. Both a student's knowledge state and a
target course are represented as low-dimensional embedding vectors and the
similarity between them is modeled by their inner product. A student's
knowledge state is implicitly computed as a linear combination of the so-called
provided knowledge component vectors of the previously-taken courses, weighted
by his/her grades in them. Though CKRM was shown to provide state-of-the-art
grade prediction accuracy, it is limited in that it assumes that all historical
courses contribute equally in estimating the student's grade in a future
course. Intuitively, students take courses from different departments, and each
course would require an acquisition of knowledge from a few other courses, with
different weights.

Motivated by the success of neural attentive networks in different
fields~\cite{he2018nais,mei2018attentive,he2017neural,bahdanau2014neural,xiao2017attentional},
in this paper, we improve upon CKRM by learning the different importance of
previously-taken courses in estimating the grade of a future course. We
leverage the recent advances in neural attentive networks to learn these
different weights, by employing both softmax and sparsemax activation functions
that output posterior probabilities, i.e., attention weights, for the prior
courses. The sparsemax function has an additional benefit of truncating the
small probability values to zero, assigning zero effect to the irrelevant prior
courses when predicting a target course's grade.

The main contributions of this work are as follows:
\begin{enumerate}
\item We propose a Neural Attentive Knowledge-based model (\method) for grade
  prediction that improves upon CKRM by employing the attention mechanism in
  neural networks to learn the different importance of the prior courses
  towards predicting the grades of target courses. To our knowledge, this is
  the first work to apply attentive neural networks to grade prediction.

\item We leverage the recent sparsemax activation function for the attention
  mechanism that produces sparse attention weights instead of soft attention
  weights.

\item We performed an extensive experimental evaluation on a real world dataset
  obtained from a large university that spans a period of 16 years and consists
  of $\sim$1.5 grades. The results show that our proposed \method\ model
  significantly improves the prediction accuracy compared to the competing
  models. In addition, the results show the effectiveness of the attention
  mechanism in learning the different importance of the previously-taken
  courses towards each target course, which can help in designing better degree
  plans and more informed course selection decisions.

\end{enumerate}

\section{Definitions and Notations}
\label{sec:notations}

Boldface uppercase and lowercase letters will be used to represent matrices and
vectors, respectively, e.g., $\G$ and $\p$. The $i$th row of matrix
$\mathbf{P}$ is represented as $\p_i^T$, and its $j$th column is represented as
$\p_j$. The entry in the $i$th row and $j$th column of matrix $\G$ is denoted
as $g_{i,j}$. A predicted value is denoted by having a hat over it (e.g.,
$\hat{g}$).

Matrix $\G$ will represent the $m \times n$ student-course grades matrix, where
$g_{s,c}$ denotes the grade that student $s$ obtained in course $c$, relative
to his/her average previous grade. Following the row-centering technique that
was first proposed by Polyzou \etal~\cite{polyzou2016grade}, we subtract each
student's grade from his/her average previous grade, since this was shown to
significantly improve the prediction accuracy of different models. As there can
be some students who achieved the same grades in all their prior courses, and
hence their relative grades will be zero, in this case, we assigned a small
value instead, i.e., 0.01. This is to prevent a prior course from not being
considered in the model computation. A student $s$ enrolls in sets of courses
in consecutive terms, numbered relative to $s$ from $1$ to the number of terms
in he/she has enrolled in the dataset. A set $\term_{s,w}$ will denote the set
of courses taken by student $s$ in term $w$.

\section{Related Work}

\subsection{Grade Prediction Methods}
\label{sec:related:grade-prediction}

Grade prediction approaches for courses not yet taken by students have been
extensively explored in the
literature~\cite{ren2017grade,ren2018ale,hu2018course,sweeney2016next,polyzou2016grade,morsy2017cumulative,elbadrawy2016domain}. In
this section, we review some research in grade prediction that is most relevant
to our work.

\subsubsection{Course-Specific Regression Models (CSR)}
\label{sec:related:csr}

A more recent and natural way to model the grade prediction problem is to model
the way the academic degree programs are structured. Each degree program would
require the student to take courses in a specific sequencing such that the
knowledge acquired in previous courses are required for the student to perform
well in future courses. Polyzou \etal~\cite{polyzou2016grade} developed
course-specific linear regression models (CSR) that build on this idea. A
student's grade in a course is estimated as a linear combination of his/her
grades in previously-taken courses, with different weights learned for each
(prior, target) course pair. For a student $s$ and a target course $j$, the
predicted grade is estimated as:
\begin{equation}
  \label{eq:csr}
  \hat{g}_{s,j} = cb_j + \sum_{i \in \mathcal{P}} w_{i, j} ~ g_{s, i},
\end{equation}
\noindent where $cb_j$ is the bias terms for course $j$, $w_{i, j}$ is the
weight of course $i$ towards predicting the grade of course $j$, $g_{s, i}$ is
the grade of student $s$ in course $i$, and $\mathcal{P}$ is the set of courses
taken by $s$ prior to taking course $j$. To achieve high prediction accuracy,
CSR requires sufficient training data for each (prior, target) pair, which can
hinder these models from good generalization.

\subsubsection{Cumulative Knowledge-based Regression Models (CKRM)}
\label{sec:related:ckrm}

Morsy \etal~\cite{morsy2017cumulative} developed Cumulative Knowledge-based
Regression Models (CKRM), which is also based on the fact that the student's
performance in a future course is based on his/her performance in the
previously-taken courses. It assumes that a space of knowledge components
exists such that each course provides a subset of these components as well as
requires the knowledge of some of these components from the student in order to
perform well in it. The student by taking a course thus acquires its knowledge
components in a way that depends on his/her grade in that course. The overall
knowledge acquired by the student after taking a set of courses is then
represented by a knowledge state vector that is computed as the sum of the
knowledge component vectors of those courses, weighted by his/her grades in
them. Let $\p_i$ denote the provided knowledge component vector for course
$i$. The knowledge state vector for student $s$ at term $t$ can be expressed as
follows:
\begin{equation}
  \label{eq:ks}
  \ks_{s,t} = \sum_{w = 1}^{t-1}  \xi(s, w, t) \sum_{i \in \term_{s, w}} \Big( g_{s,i} ~ \p_i \Big),
\end{equation}
\noindent where $g_{s,i}$ is the grade that student $s$ obtained on course
$i$, and $\xi(s, w, t)$ is a time-based exponential decaying function
designed to de-emphasize courses that were taken a long time ago.

Given the student's knowledge state vector prior to taking a course and that
course's required knowledge component vector, denoted as $r_j$, CKRM estimates
the student's expected grade in that course as the inner product of these two
vectors, i.e.,
\begin{align}
  \label{eq:ckrm}
  \hat{g}_{s,j} = cb_j + \ks_{s,t}^T ~ \req_j,
\end{align}
\noindent where $cb_j$ is as defined in Eq.~\ref{eq:csr}, and $\ks_{s,t}$ is
the corresponding knowledge state vector. These course-specific linear models
are estimated from the historical grade data and can be considered as capturing
and weighting the knowledge components that a student needs to have accumulated
in order to perform well in a course.


\subsection{Neural Attentive Models}


Our work relies on the attention mechanism, which has been recently introduced
in neural networks and was shown to improve the performance of different models
and give better explanations to the importance of different objects towards a
target
object~\cite{he2017neural,xiao2017attentional,he2018nais,chen2017attentive}. Our
work leverages several advances in this area. The most commonly-used activation
function for the attention mechanism is the softmax function, which is easily
differentiable and gives soft posterior probabilities that normalize to 1. A
major disadvantage of the softmax function is that it assumes that each object
contributes to the compressed representation, which may not always hold in some
domains. To solve this, we need to output sparse posterior probabilities and
assign zero to the irrelevant objects. Martins \etal~\cite{martins2016softmax}
proposed the sparsemax activation function, which has the benefit of assigning
zero probabilities to some output variables that may not be relevant for making
a decision. This is done by defining a threshold, below which small probability
values are truncated to zero. We also leverage the controllable sparsemax
activation function recently proposed by Laha \etal~\cite{laha2018controllable}
that controls the desired degree of sparsity in the output probabilities. This
is done by adding an L2 regularization term that is to be maximized in the loss
function. This will potentially encourage larger probability values for some
objects, moving the rest to zero.

\begin{table*}[ht]
  \renewcommand{\arraystretch}{1.3}
  \caption{Sample of prior and target courses for a Computer Science student at University X.}
  \begin{center}
  \begin{scriptsize}
  \begin{tabular}{lc}
    \toprule
    \multicolumn{1}{c}{Prior Courses} & Target Course \\
    \midrule
    \multirow{3}{*}{\parbox{13cm}{Calculus I, Beginning German, Operating
    Systems, Intermediate German I, University Writing, Introductory
    Physics, Peotics in Film,
    Program Design \& Development, Philosophy, Linear Algebra,
                      Internet Programming, Stone Tools to Steam Engines, Advanced Programming
                      Principles, Computer Networks}} & Intermediate German II
  \\
    \cmidrule{2-2}
                                        & Probability \& Statistics \\
    \cmidrule{2-2}
                  & Algorithms \& Data Structures \\
    \bottomrule
  \end{tabular}
  \label{tbl:sample-courses}
  \end{scriptsize}
\end{center}

\end{table*}

\section{Proposed Model} 

\subsection{Motivation}
\label{sec:methods:motivation}

Consider a sample student who is declared in a Computer Science major and is in
his/her second or third year in college. Table~\ref{tbl:sample-courses} shows
the set of prior courses that this student has already take and the set of
courses that this student is planning on taking the next term. With CKRM
(Section~\ref{sec:related:ckrm}), all these prior courses would contribute
equally to predicting the grade of each target course. However, we can see
that, intuitively, from the courses' names, there are courses that are strongly
related to each target course and other courses that are irrelevant to it. For
instance, it is reasonable to expect that the Intermediate German II course is
more related to the Intermediate German I course than any of the other courses
that the student has already taken. Along the same lines, we expect that the
Algorithms and Data Structures course is more related to other Computer Science
courses, such as the Advanced Programming Principles and the Program Design and
Development courses. Assuming equal contribution among these prior courses can
hinder the grade prediction model from accurately learning the course
representations, and hence lead to poor predictions.

\subsection{Overview}

In this work, we present our Neural Attentive Knowledge-based model, \method,
which predicts a students' grades in future courses by employing an attention
mechanism on the prior courses. We use CKRM as the underlying model (see
Section~\ref{sec:related:ckrm}).

\subsection{Attention-based Pooling Layer for Prior Courses}
\label{sec:methods:attention-prior}

In order to learn the different contributions of the prior courses in
estimating the student's grade in a future course, we can employ the CSR
technique (see Section~\ref{sec:related:csr}) that learns the importance of
each prior course in estimating the grade of each future course. Thus, we would
estimate a knowledge state vector for each target course $j$, using the
following equation:
\begin{equation}
  \label{eq:nak}
  \ks_{s,t,j} = \sum_{w = 1}^{t-1} \sum_{i \in \term_w} \Big( a_{i, j} ~ g_{s,i} ~ \p_{i} \Big),
\end{equation}
\noindent where $a_{i, j}$ is a learnable parameter that denotes the attention
weight of course $i$ in contributing to student $s$'s knowledge state when
predicting $s$'s grade in course $j$. Note that we have removed the
time-decaying function $\xi(s, w, t)$ that was used in CKRM (see
Eq.~\ref{eq:ks}), since it would be implicitly included in the attention
weights. However, this solution requires sufficient training data for each
$(i, j)$ pair in order to be considered an accurate estimation.

In order to be able to have accurate attention weights between all pairs of
prior and target courses, even the ones that do not appear together in the
training data, we propose to use the attention mechanism that was recently used
in neural networks~\cite{bahdanau2014neural,vaswani2017attention}. The main
idea is to estimate the attention weight $a_{i, j}$ from the embedding vectors
for courses $i$ and $j$.

In order to compute the similarity between the embeddings of prior course $i$
and target course $j$, we use a single-layer perceptron as follows:
\begin{equation}
  \label{eq:slp-prior}
  z_{i, j} = {\mathbf{h}}^T \textrm{RELU}(\mathbf{W} (\q_i \odot \req_j) + \mathbf{b}),
\end{equation}
\noindent where $\q_i = g_{s,i} \p_i$ denotes the embedding of the prior course
$i$, weighted by the student's grade in it, $\odot$ denotes the Hadamard
product, and $\mathbf{W} \in \mathcal{R}^{l \times d}$ and
$\mathbf{b} \in \mathcal{R}^{l}$ denote the weight matrix and bias vector that
project the input into a hidden layer, respectively, and
$\mathbf{h} \in \mathcal{R}^{l}$ is a vector that projects the hidden layer
into an output attention weight, where $d$ and $l$ denote the number of
dimensions of the embedding vectors and attention network, respectively. RELU
denotes the Rectified Linear Unit activation function that is usually used in
neural attentive networks.

\subsubsection{Softmax Activation Function}
\label{sec:methods:attention-prior:softmax}

The most common activation function used for computing these attention weights
is the softmax function~\cite{vaswani2017attention}. Given a vector of real
weights $\mathbf{z}$, the softmax activation function converts it to a
probability distribution, which is computed component-wise as follows:
\begin{equation}
  \label{eq:softmax-act}
  \textrm{softmax}_i(\mathbf{z}) = \frac{\exp(z_i)}{\sum_j \exp(z_j)}.
\end{equation}
\noindent We will refer to the \method\ model that uses the softmax activation
function as {\bf \method(soft)}.

\subsubsection{Sparsemax Activation Function}
\label{sec:methods:attention-prior:sparsemax}

Although the softmax activation function has been used to design attention
mechanisms in many
domains~\cite{parikh2016decomposable,bahdanau2014neural,he2017neural,mei2018attentive,he2018nais},
we believe that using it for grade prediction is not optimal. Since a student
enrolls in several courses, and each course requires knowledge from one or a
few other courses, we hypothesize that some of the prior courses should have no
effect, i.e., zero attention, towards predicting a target course's grade. We
thus leverage a recent advance, the sparsemax activation
function~\cite{martins2016softmax}, to learn sparse attention weights. The idea
is to define a threshold, below which small probability values are truncated to
zero. Let
$\triangle^{K-1} := \{ \mathbf{p} \in \mathbb{R}^K | \mathbf{1}^T\mathbf{p} =
1, \mathbf{p} \ge \mathbf{0}\}$ be the $(K-1)$-dimensional simplex. The
sparsemax activation function tries to solve the following equation:
\begin{equation}
  \textrm{sparsemax}(\mathbf{z}) = \underset{\mathbf{p} \in
    \triangle^{K-1}}{\textrm{argmin}} ~ \| \mathbf{p} - \mathbf{z} \|^2,
\end{equation}
\noindent which, in other words, returns the Euclidean projection of the input
vector $\mathbf{z}$ onto the probability simplex.

In order to obtain different degrees of sparsity in the attention weights, Laha
\etal~\cite{laha2018controllable} developed a generic probability mapping
function for the sparsemax activation function, which they called
{\bf sparsegen}, and is computed as follows:
\begin{equation}
  \textrm{sparsegen}(\mathbf{z}; \gamma) = \textrm{argmin} ~ \| \mathbf{p} -
  \mathbf{z} \|^2 - \gamma \| \mathbf{p} \|^2,
\end{equation}
\noindent where $\gamma < 1$ controls the L2 regularization strength of
$\mathbf{p}$. An equivalent formulation for sparsegen was formed as:
\begin{equation}
  \label{eq:sparsegen-act}
  \textrm{sparsegen}(\mathbf{z}; \gamma) = \textrm{sparsemax} \big( \frac{\mathbf{z}}{1-\gamma} \big),
\end{equation}
\noindent which, in other words, applies a temperature parameter to the
original sparsemax function. Varying this temperature parameter can change the
degree of sparsity in the output variables. By setting $\gamma = 0$, sparsegen
becomes equivalent to sparsemax. We will refer to the \method\ model that uses
the sparsegen activation function as {\bf \method(sparse)}.

\subsection{Prediction}

\method\ then predicts the grade for student $s$ in course $j$ that he/she
takes at term $t$ as:
\begin{equation}
  \label{eq:cnak}
  \hat{g}_{s,j} = cb_j + \ks_{s,t,j}^T ~ \req_j.
\end{equation}

\subsection{Optimization}

We use the mean squared error (MSE) loss function to estimate
the parameters of \method. We minimize the following regularized RMSE loss:
\begin{equation}
  \label{eq:opt}
  L = -\frac{1}{2N} \sum_{{s,c} \in \G} {(g_{s,c} - \hat{g}_{s,c})}^2 + \alpha||\Theta||^2\\,
\end{equation}
where $N$ is the number of grades in $\G$. The hyper-parameter $\alpha$
controls the strength of L2 regularization to prevent overfitting, and
$\Theta = \{\{\mathbf{cb}\}, \{\p_i\}, \{\req_i\},
\mathbf{W}, \mathbf{b}, \mathbf{h}\}$ denotes all trainable parameters of
\method.

The optimization problem is solved using AdaGrad
algorithm~\cite{duchi2011adaptive}, which applies an adaptive learning rate for
each parameter. It randomly draws mini-batches of a given size from the
training data and updates the related model parameters.

\section{Evaluation Methodology}

\subsection{Dataset}

The data used in our experiments was obtained from the University of Minnesota
(UMN), which includes 96 majors from 10 different colleges, and spans the years
$2002$ to $2017$. At the University, the letter grading system used is A--F,
which is converted to the 4--0 scale using the standard letter grade to GPA
conversion. We removed any grades that were taken as pass/fail. The final
dataset includes $\sim 54,000$ students, $5,800$ courses, and $1,450,000$
grades in total.

\begin{table*}[t]
  \caption{Comparison between the baseline and proposed models.}
  \centering
  \begin{scriptsize}
    \begin{tabular}{lrrrrrllll}
      \toprule
      \multicolumn{1}{c}{Model} & \multicolumn{5}{c}{Parameters} &
                                                                   \multicolumn{1}{c}{RMSE}
      & \multicolumn{1}{c}{PTA0} & \multicolumn{1}{c}{PTA1} &
                                                              \multicolumn{1}{c}{PTA2}
      \\
      \midrule
      MF & 16 & 1E-04 & 1E-02 & -- & -- & 0.724 & 25.7 & 58.6 & 79.5 \\
      KRM(sum) & 32 & 1E-07 & 7E-04 & 0.3 & -- & 0.584 & 32.6 & 70.1 & 87.7 \\
      KRM(avg) & 32 & 1E-07 & 7E-04 & 0.0 & -- &  0.584 & 34.9 & 70.6 & 87.7 \\
      \hline
      \method(soft) & 32 & 1E-07 & 7E-04 & 3 & -- & 0.589 (-0.9\%) & \ul{35.3}\dag\
                                                                     (1.1\%) &
                                                                               71.8 (1.7\%)
                                             & 88.0\dag\ (0.3\%) \\
      \method(sparse) & 32 & 1E-07 & 7E-04 & 4 & 0.5 & \ul{0.574}\dag\ddag\
                                                       (1.7\%) & \ul{35.3}\dag\ (1.1\%) &
                                                                              \ul{72.1} (2.1\%)
                                             & \ul{88.7}\dag\ (1.1\%) \\
      \bottomrule
      \end{tabular}
        \begin{tablenotes}[paraleft]
        \item[] The Parameters columns denote the following model parameters
          that were selected: for MF, the parameters are: the number of latent
          dimensions, the L2 regularization parameter, and the learning rate;
          for KRM(sum) and KRM(avg), the parameters are: the embedding size for
          courses, the L2 regularization parameter, the learning rate, and the
          time-decaying parameter $\lambda$; for \method, the parameters are:
          the embedding size for courses, the L2 regularization parameter, the
          learning rate, and the number of latent dimensions for the MLP
          attention mechanism; and for \method(sparse), the last parameter
          denotes the L2 regularization parameter $\gamma$ for the sparsegen
          activation function. Underlined entries represent the best
          performance in each metric. The \dag\ and \ddag\ symbols are used to
          denote results that are statistically significant over the best
          performing baseline metric, and \method(soft), respectively, using
          the Student's paired $t$-test with a $p$-level < 0.1. Numbers in
          parentheses denote the percentage of improvement over the best
          baseline value in each metric.
        \end{tablenotes}
      \end{scriptsize}
      \label{tbl:main-res}
\end{table*}

\subsection{Generating Training, Validation and Test Sets}

At UMN, there are three terms, Fall, Summer and Spring. We used the data from
$2002$ to Spring $2015$ (inclusive) as the training set, the data from Spring
$2016$ to Fall $2016$ (inclusive) as the validation set, and the data from
Summer $2016$ to Summer $2017$ (inclusive) as the test set. For a target course
taken by a student to be predicted, that student must have taken at least four
courses prior to the target course, in order to have sufficient data to compute
the student's knowledge state vector. We excluded any courses that do not
appear in the training set from the validation and test sets.

\subsection{Comparison Methods}

We compared the performance of our \method\ model against the following grade
prediction approaches:
\begin{enumerate}
\item {\bf{Matrix Factorization (MF):}} This approach predicts the grade for
  student $s$ in course $i$ as:
  \begin{equation}
    \label{eq:mf}
  \hat{g}_{s,i} = \mu + sb_s + cb_i + \mathbf{u}_s^T ~ \mathbf{v}_i,
\end{equation}
\noindent where $\mu$, $sb_s$ and $cb_i$ are the global, student and course
bias terms, respectively, and $\mathbf{u}_s$ and $\mathbf{v}_i$ are the student
and course latent vectors, respectively. We used the squared loss function with
L2 regularization to estimate this
model.

\item {\bf KRM(sum):} This is CKRM the method described in
  Section~\ref{sec:related:ckrm}.

\item {\bf KRM(avg):} This is similar to the KRM(sum) method, except that the
  prior courses' embeddings are aggregated with mean pooling instead of
  summation. It was shown in later studies, e.g.~\cite{ren2018ale}, that it
  performs better than KRM(sum).
\end{enumerate}

We implemented KRM(sum) and KRM(avg) with a neural network architecture and
optimization similar to that of \method.

\subsection{Model Selection}

We performed an extensive search on the parameters of the proposed and baseline
models to find the set of parameters that gives us the best performance for
each model.

For all proposed and competing models, the following parameters were used. The
number of latent dimensions for course embeddings was chosen from the set of
values: \{8, 16, 32\}. The L2 regularization parameter was chosen from the
values: \{1e-5, 1e-7, 1e-3\}. Finally, the learning rate was chosen from the
values: \{0.0007, 0.001, 0.003, 0.005, 0.007\}. For the proposed \method\
models, the number of latent dimensions for the MLP attention mechanism was
selected in the range [1, 4]. For KRM(sum) and KRM(avg), the time-decaying
parameter $\lambda$ was chosen from the set of values: \{0, 0.3, 0.5, 0.7,
1.0\}.

The training set was used for estimating the models, whereas the validation set
was used to select the best performing parameters in terms of the overall MSE
of the validation set.

\subsection{Evaluation Methodology and Metrics}

The grading system used by the University uses a 12 letter grade system (i.e.,
A, A-, B+, $\ldots$ F). We will refer to the difference between two successive
letter grades (e.g., B+ vs B) as a \emph{tick}. We converted the predicted
grades into their closest letter grades. We assessed the performance of the
different approaches based on the Root Mean Squared Error (RMSE) as well as how
many ticks away the predicted grade is from the actual grade, which is referred
to as {\em Percentage of Tick Accuracy}, or PTA. We computed the percentage of
grades predicted with no error (zero tick), within one tick, and within two
ticks, which will be referred to as PTA0, PTA1, and PTA2, respectively.


\section{Experimental Results}


\subsection{Performance of the Proposed Models}

Table~\ref{tbl:main-res} shows the performance of our proposed models. Using
the sparsegen activation function instead of the softmax activation function
improves the prediction accuracy, with a statistically significant improvement.
This shows that using the sparsegen activation function to output sparse
attention weights for the prior courses achieves better prediction accuracy
than producing soft probabilities for all of them. This is expected, since the
student's prior courses may not be all relevant to the target course, as
illustrated in Table~\ref{tbl:sample-courses}.

\subsection{Performance against Competing Methods}

Table~\ref{tbl:main-res} also shows the performance of the competing
models. Among the baseline methods, both KRM(sum) and KRM(avg) outperform
MF. KRM(avg) outperforms KRM(sum) in PTA0 and PTA1. Both \method(soft) and
\method(sparse) outperform all baseline methods. Even though the RMSE results
of \method(soft) is worse than these of the KRM variants, it achieved
$\sim$1\%, $\sim$2\% and 0.5\% more accurate predictions within no, one, and
two tick errors, respectively. Among all baseline and proposed methods, our
\method(sparse) model outperforms all baseline methods significantly, with
achieving $\sim$2\% lower RMSE, and $\sim$1\% more accurate predictions within
two ticks than KRM(avg). This shows that using the attention-based pooling
layer on the prior courses to accumulate them can better predict the grades of
students in their future courses.

\begin{table*}[ht]
  \renewcommand{\arraystretch}{1.3}
  \caption{The attention weights of the prior courses with each target course
    learned by \method(sparse) for the sample student from
    Table~\ref{tbl:sample-courses}.}
  \begin{center}
  \begin{scriptsize}
  \begin{tabular}{p{13cm}c}
    \toprule
    \multicolumn{1}{c}{Prior Courses} & Target Course \\
    \midrule
    Intermediate German I: \red{0.6980}, University Writing: \red{0.3020} & Intermediate German II \\
    \midrule
    Calculus I: \red{0.4737}, Physics: \red{0.3794}, Program Design \&
                                                                       Development: \red{0.0717},
                                                                       Operating
                                                                       Systems: \red{0.0497},
                                                                       Computer
                                                                       Networks: \red{0.0255} & Probability \&
                                                                                 Statistics
    \\
    \midrule
    Operating Systems: \red{0.2927},
    Advanced Programming Principles: \red{0.2582}, Linear Algebra:
    \red{0.2313}, Physics: \red{0.2178}  & Algorithms \& Data Structures \\
    \bottomrule
  \end{tabular}
  \begin{tablenotes}[paraleft]
    \item[] Prior courses are sorted in non-increasing order w.r.t. to their
      attention weights with each target courses for clarity purposes.
  \end{tablenotes}
  \label{tbl:nak-attn-weights}
  \end{scriptsize}
\end{center}

\end{table*}

\subsection{Qualitative Analysis on the Prior Courses Attention Weights}

Recall the motivational example for the Computer Science student, discussed in
Section~\ref{sec:methods:motivation}. This student had a set of prior courses
and three target courses that we would like to predict his/her grades in (See
Table~\ref{tbl:sample-courses}). Using KRM(sum) or KRM(avg), all the prior
courses would contribute equally to the prediction of each target course. Using
our proposed \method(sparse) model, the attention weights for the prior courses
with each target course are shown in
Table~\ref{tbl:nak-attn-weights}\footnote{These results were obtained by
  learning \method\ models to estimate the actual grades and not the
  row-centered grades. This allowed us to get more interpretable
  results.}.

We can see that, using the sparsegen activation function, only a few prior
courses are selected with non-zero attention weights, which are the most
relevant to each target course.

For the Intermediate German II course, we can see that the student's grade in
it is most affected by two courses: the Intermediate German I course, and the
University Writing course. The Intermediate German I course is listed as a
pre-requisite course for the Intermediate German II course. Though the
University Writing course is not listed as a pre-requisite course, after
further analysis, we found out that the Intermediate German II course requires
process-writing essays and are considered part of the grading system. Though
the German courses are not part of the student's degree program, and are taken
by a small percentage of Computer Science students, our \method\ model was able
to learn accurate attention weights for them.

The other two target courses, Probability and Statistics, and Algorithms and
Data Structures, have totally different prior courses with the largest
attention weights, which are more related to them.

These results illustrate that our proposed \method\ model was able to uncover
the listed as well as the hidden/informal pre-requisite courses without any
supervision given to the model.

\section{Conclusion}

In this work, we presented a method to improve the grade prediction accuracy,
by learning the weights of the prior courses towards predicting the grade of
each target course. To this end, we employed the attention mechanism on the
prior courses that learns the different contributions of these courses towards
each target course. We employed both a softmax and a sparsemax activation
function that produce soft and sparse attention weights, respectively. The
proposed models are able to capture the listed as well as the hidden
pre-requisite courses for the target courses, which can be better used to
design better degree plans. Our experiments showed that our models
significantly outperformed the competing methods, indicating the value of the
attention mechanism on the prior courses.


\section*{Acknowledgement}
This work was supported in part by NSF (1447788, 1704074, 1757916, 1834251),
Army Research Office (W911NF1810344), Intel Corp, and the Digital Technology
Center at the University of Minnesota. Access to research and computing
facilities was provided by the Digital Technology Center and the Minnesota
Supercomputing Institute, \url{http://www.msi.umn.edu}.

\bibliographystyle{abbrv}
\bibliography{refs}

\balancecolumns

\end{document}